\documentclass[twoside,11pt]{article}

\usepackage{xspace}
\newcommand{\mbart}{{\tt mBART}\xspace}
\newcommand{\mtfive}{{\tt MT5}\xspace}
\newcommand{\sage}{{\tt SAGE}\xspace} 
\newcommand{\attr}{{\textsuperscript{\tt[*A*]}}}

\newcommand{\msage}{{\tt mSAGE}\xspace}

\newcommand{\asin}{{product}\xspace}
\newcommand{\asins}{{products}\xspace}
\newcommand{\sabex}{{\tt MLC}\xspace}
\newcommand{\atlas}{{\tt mBERT}\xspace}  
\newcommand{\model}{{\tt MODEL}\xspace}
\newcommand{\maveqa}{{\tt MAVEQA}\xspace}

\newcommand{\testdata}{{\tt BPD\textsubscript{[500]}}}
\newcommand{\fulldata}{{\tt BPD}}
\usepackage{todonotes}
\usepackage[notheorems]{nikolako-math}
\usepackage{varwidth}
\usepackage{algorithm}
\usepackage{algpseudocode}
\usepackage{tabularx}
\usepackage{multicol}
\usepackage{booktabs} 
\algblock{ParFor}{EndParFor}
\algnewcommand\algorithmicparfor{\textbf{parfor}}
\algnewcommand\algorithmicpardo{\textbf{do}}
\algnewcommand\algorithmicendparfor{\textbf{end\ parfor}}
\algrenewtext{ParFor}[1]{\algorithmicparfor\ #1\ \algorithmicpardo}
\algrenewtext{EndParFor}{\algorithmicendparfor}

\newif\ifblackandwhite

\usepackage{etoolbox}
\usepackage{pdflscape}
\usepackage{colortbl}%
\newcommand{\myrowcolour}{\rowcolor{white!89.803921568627459!black}}
\usepackage[caption=false]{subfig}

\usepackage[flushleft]{threeparttable}

\definecolor{Maroon}{cmyk}{0,1,1,0.5}

\ifblackandwhite

\newcommand{\highest}[1]{\textbf{#1}}
\else

\newcommand{\highest}[1]{\textcolor{Maroon}{\textbf{#1}}}%
\fi

\usepackage{pgfplots}
\pgfplotsset{compat=newest}
\usepgfplotslibrary{groupplots}
\usepgfplotslibrary{polar}
\usepgfplotslibrary{statistics}
\usetikzlibrary{positioning,chains,fit,shapes,calc}

\definecolor{mycolor1}{rgb}{0.9719668012419661,0.4636196661831345,0.4272112904308525}
\definecolor{mycolor2}{rgb}{0.8273618134500182,0.5740127684739248,0.0}
\definecolor{mycolor3}{rgb}{0.576235910262533,0.6647921100239956,0.0}
\definecolor{mycolor4}{rgb}{0.0,0.7281510724816163,0.22080268587160473}
\definecolor{mycolor5}{rgb}{0.0,0.7553354789496858,0.6252760328031264}
\definecolor{mycolor6}{rgb}{0.0,0.7248676308880103,0.8911735834721597}
\definecolor{mycolor7}{rgb}{0.3804515572007965,0.6115286754014934,1.0}
\definecolor{mycolor8}{rgb}{0.857783940581795,0.44743713611909713,0.9846138428363592}
\definecolor{mycolor9}{rgb}{1.0,0.38135337340068687,0.7651117134811273}

\xdefinecolor{GopherMaroon}{HTML}{7A0019}
\xdefinecolor{GopherGold}{HTML}{FFCC33}
\xdefinecolor{GopherLightGold}{HTML}{FFDE7A}
\xdefinecolor{GopherDarkMaroon}{HTML}{5B0013}

\definecolor{Maroon}{RGB}{238,83,34}
\definecolor{Maroon}{RGB}{25,89,121}
\definecolor{bgorange}{RGB}{238,83,0}
\definecolor{bgorangelight}{RGB}{255,173,115}
\definecolor{bggreen}{RGB}{237,255,178}
\definecolor{bgblue}{RGB}{128,255,204}
\definecolor{Maroon}{RGB}{122,0,25}
\definecolor{Gold}{RGB}{255,204,51} 

\usetikzlibrary{shapes}
\definecolor{pinegreen}{cmyk}{0.92,0,0.59,0.25}
\definecolor{royalblue}{cmyk}{1,0.50,0,0}
\definecolor{lavander}{cmyk}{0,0.48,0,0}
\definecolor{violet}{cmyk}{0.79,0.88,0,0}
\tikzstyle{citems}=[circle, draw, thin,fill=Gold, scale=0.8]
\tikzstyle{cusers}=[rectangle, draw, thin,fill=Maroon, scale=0.8]
\tikzstyle{cusers2}=[rectangle, draw, thin,fill=white, scale=0.8]
\tikzstyle{cred}=[circle, draw, thin,fill=Maroon, scale=0.8]
\tikzstyle{cgreen}=[rectangle, draw, thin,fill=lavander, scale=0.8]
\tikzstyle{rpath}=[ultra thick, Maroon, opacity=0.8]
\tikzstyle{gpath}=[ultra thick, royalblue, opacity=0.5]

\usepackage{wrapfig,lipsum,booktabs}
\usepackage{tcolorbox}
\usepackage{adjustbox}

\makeatletter
\global\let\tikz@ensure@dollar@catcode=\relax
\makeatother

\usepackage{blindtext}

\usepackage[abbrvbib, preprint]{jmlr2e}

\usepackage{lastpage}
\jmlrheading{23}{2023}{1-\pageref{LastPage}}{6/23; Revised 5/22}{9/22}{21-0000}{Nikolakopoulos \textit{et al}}

\ShortHeadings{Nikolakopoulos et al.}{SAGE}
\firstpageno{1}

\begin{document}

\title{SAGE: Structured Attribute Value Generation for Billion-Scale Product Catalogs}

\author{\name Athanasios N. Nikolakopoulos \email athanani@amazon.com \\
       \addr Amazon Catalog AI\\
       550 Terry Ave N, Seattle, WA 98109, USA
       \AND
       \name Swati Kaul \email kauswati@amazon.com \\
       \addr Amazon Catalog AI\\
       550 Terry Ave N, Seattle, WA 98109, USA
       \AND
       \name Siva Karthik Gade~\thanks{Work done while the author was in Amazon} \email sivakarthik.gade@gmail.com\\
       \addr Meta\\
		1101 Dexter Ave N, Seattle, WA 98109, USA
       \AND
       \name Bella Dubrov \email belladub@amazon.com \\
       \addr Amazon Catalog AI\\
       550 Terry Ave N, Seattle, WA 98109, USA
       \AND
       \name Umit Batur \email baturab@amazon.com \\
       \addr Amazon Catalog AI\\
       550 Terry Ave N, Seattle, WA 98109, USA
       \AND
       \name Suleiman Ali Khan \email suleimkh@amazon.com \\
       \addr Amazon Catalog AI\\
       550 Terry Ave N, Seattle, WA 98109, USA
   }

\editor{}

\maketitle

\begin{abstract}
We introduce \sage; a Generative LLM for inferring attribute values for \asins across world-wide e-Commerce catalogs. We introduce a novel formulation of the attribute-value prediction problem as a Seq2Seq summarization task, across languages, product types and target attributes. Our novel modeling approach lifts the restriction of predicting attribute values within a pre-specified set of  choices, as well as, the requirement that the sought attribute values need to be explicitly mentioned in the text.
\sage can infer attribute values even when such values are mentioned implicitly using periphrastic language, or not-at-all---as is the case for common-sense defaults. 
Additionally, \sage is capable of predicting whether an attribute is \textit{inapplicable} for the \asin at hand, or \textit{non-obtainable} from the available information. \sage is the first method able to tackle all aspects of the attribute-value-prediction task as they arise in practical settings in e-Commerce catalogs. A comprehensive set of experiments demonstrates the effectiveness of the proposed approach, as well as, its superiority against state-of-the-art competing alternatives. Moreover, our experiments  highlight \sage's ability to tackle the task of predicting attribute values in zero-shot setting; thereby, opening up opportunities for significantly reducing the overall number of labeled examples required for training.  
\end{abstract}


\section{Introduction}

Product listings on E-commerce catalogs comprise of unstructured text including the title, bullet points and description alongside possible images related to the product. Products are organized in product-types, and each product-type (PT) is associated with a list of relevant \textit{attributes} that highlight important aspects of the product, such as \textit{material}, \textit{color}, \textit{shape}, etc. Customers rely on such structured attribute information to search, browse, compare, and ultimately decide which products to purchase. Predicting attribute values from unstructured product text is a fundamental challenge for world-wide e-Commerce catalogs such as Amazon, Walmart and AliBaba. Rich meta data are essential for deep understanding of the products, which in turn, is valuable for critical downstream applications, such as  recommendations, search, question answering; as well as, for providing an enhanced customer experience. 

The importance of the attribute-value-prediction task has sparked a lot of research over the recent years in both academia and industry~\citep{GhaniPLKF06,ProbstGKFL07,CarmelLM18,RezkANZ19,ZhaoGS19,EXACT}. First attempts to tackle the problem relied on rule-based methods~\citep{chiticariu-etal-2010-domain,VANDIC2012425}, 
which relied on regular expressions relying on domain-specific knowledge, 
but with the advent of transformers~\citep{attention} and other advances in Natural Language Understanding, the focus quickly shifted towards more general ML solutions. An important family of methods cast the underlying task as an Named Entity Recognition (NER) problem~\citep{PutthividhyaH11,More16,https://doi.org/10.48550/arxiv.2106.01223,Nadeau2007ASO} and build extraction models to identify the attribute values within the input text. Another line of research employs sequence tagging models for attribute value extraction~\citep{10.1145/3219819.3219839,xu-acl2019-scaling,10.1145/3394486.3403047,Yan2021} .  More recently, Google introduced \maveqa\citep{MAVE}; an extraction-based method which casts attribute value prediction as a question-answering problem, showing promising results.

However, the aforementioned methods are fundamentally  limited to extracting values that are explicitly mentioned in the text. As such, they are destined to meet an invisible recall barrier above which they are inherently unable to reach. \textit{What happens if the sought attribute value is not present in the text?} 

\begin{figure}
	\centering
	\includegraphics[width=0.92\linewidth]{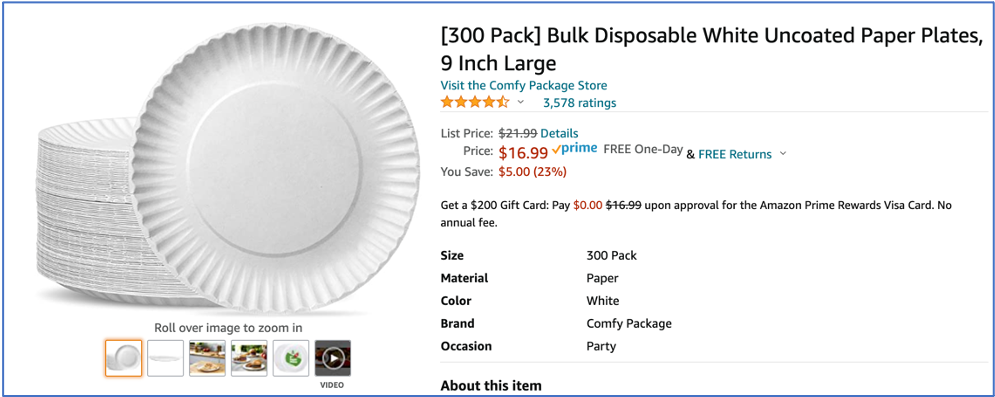}
	\caption{Example of Attribute Value Generation. The text or image does not specify if the paper plate is dishwasher safe or otherwise, however, implicit language and knowledge relationships between paper, water and dishwasher can help identify the correct attribute value.}
	\label{fig:gen_example}
\end{figure}

For approaches predicated on attribute value extraction this is an insurmountable hurdle. It prevents predicting values that are unmentioned defaults (such as ``unflavored''), as well as, ones that are inferable but not extractable (such as an \textit{item\_shape} being ``round'' when the diameter is mentioned). 
Similarly, boolean attributes are often skipped by the sellers when they are false, or when common sense would make their value immediately obvious to the customer (e.g., sellers do not say that paper plates are not dishwasher-safe Fig.~\ref{fig:gen_example}). Extraction-based methods, when confronted with such cases will inevitably return empty-handed. Employing narrow-scope multi-label classifiers could circumvent this issue, from a technical point of view, however, undertaking such a task at catalog-wide scale would lead to significant challenges, such as curating valid attribute-value lists across hundreds of thousands of Product-type-Attribute-Country (PAC) scopes,  dealing with output values that change over time (e.g., attributes like \textit{style}, \textit{theme} etc are ever-evolving in practice), as well as, maintaining and updating thousands of models in production.  

\subsection{Our Contribution} Motivated by the above, in this work we attempt to break the glass ceiling imposed by extraction-based methodologies.  We propose a novel formulation for the attribute value prediction problem, and we introduce \sage; a multi-lingual transformer-based generative Seq2Seq model able to tackle the problem of attribute-value prediction across thousands of PAC scopes.  The most important novelties of \sage in context to prior art are listed below:
\begin{enumerate}
	\item \sage generated attribute values are \emph{not constrained to exist in the input}. This is a crucial property of \sage that sets it apart from most prior solutions to the attribute-value-prediction problem.  It enables \sage  to generate correct attribute values relying on non-trivial associations within the text, implicit language, as well as, to predict common-sense default values (e.g., \textit{is\_electric} should be `False' for manual toothbrushes, even when the seller does not explicitly mention it in the text). It also enables \sage to readily extend to other sources of information besides text without redesigning the problem formulation from scratch (e.g., adding images, reviews, Q\&A, etc can be done in a straightforward manner, without special modifications of the core problem formulation). Importantly,  this property allows  \sage, to tackle cases SOTA extraction-based models are fundamentally unable to address.
	\item \sage has the significant benefit of alleviating the burden of curating customized transformations of the input, or elaborate post-filtering logic, in anticipation of corner cases throughout e-Commerce catalogs.  Such interventions are an inescapable reality to any extraction-based method that aims to solve the attribute-value-prediction task across wide range of attributes. 
	\begin{itemize}
		\item For example, certain attributes can be constrained to have only a desired set of values. Let's consider the attribute \textit{water\_resistant\_level}: in many catalogs the attribute is allowed to take only the values: `water\_resistant',  `not\_water\_resistant', and `waterproof'. However the \asin text may contain terms like `water-resistance', `resistance to water', or other periphrastic descriptions of the sought attribute value.  
	\end{itemize} 
	Curating lists of synonyms for all potential attribute values across multiple languages, and writing custom logic to filter model predictions is clearly not a scalable solution for the problem. \sage with its innate capability to go beyond extractions, can learn to predict attribute values in the expected \textit{normalized} format, thereby, offering a natural and scalable solution to this problem.  
	\item \sage is able to handle \textit{multi-valued attributes} in a seamless manner; a property particularly useful for many categories of attributes in e-Commerce catalogs, like \textit{occasion\_type}, \textit{recommended\_uses\_for\_product} etc, for which many attribute values might be relevant to the customers. 
	\item \sage has the ability to make attribute-value predictions in \textit{zero-shot mode}. Using its general understanding of language and attribute domain knowledge, \sage can make predictions even for PAC scopes that it has not been explicitly trained on. This is a very useful feature that fuels the fast and economical (in terms of training labels) expansion of the model across catalog scopes of interest. It also allows the model to better handle new or unseen scopes, thus, adapting to the realities of ever-changing e-Commerce product catalogs, in an efficient manner.
	\item \sage is also able to assess attribute \textit{applicability} for the \asin at hand, as well as, the \textit{obtainability} of the target attribute value based on available information. This renders \sage the first method to be able to tackle all aspects of the problem of Catalog data completeness as they arise in practical settings. 
\end{enumerate}

 We conduct an extensive set of experiments which showcase the potential of the proposed methodology, in supervised, zero-shot and multi-modal settings; \sage achieves an average  \textbf{Recall of 84.86\% at Precision 96\% or above} across thousands of PAC scopes; significantly outperforming  existing baseline extraction-based approaches, as well as, narrow-scope classifiers.

\section{ SAGE }
\subsection{Attribute-Value-Prediction Task}
Let $\mathcal{A}$ a set of attributes, and $\mathcal{X}$ be a set of \asins. Each \asin $x \in \mathcal{X}$ can be thought-of as a textual representation of a product comprising relevant information about the \asin; e.g., its title, bullet-points, description, the product-type it belongs to, and the country that offers it. We set forth the following problem formulation for the attribute-value-prediction task:    

\begin{tcolorbox}
	Given the product representation $x \in \mathcal{X}$, and a target attribute $a \in \mathcal{A}$, learn a function 
	\begin{displaymath}
		f: \mathcal{A}\times\mathcal{X} \to \mathfrak{P}\left(\mathcal{V}\right)
	\end{displaymath}where $\mathfrak{P}\left(\mathcal{V}\right)$ is the powerset of all possible attribute values, $\mathcal{V}$.
\end{tcolorbox}

\subsection{SAGE Model}

\begin{figure}
    \centering
    \includegraphics[width=0.998\linewidth]{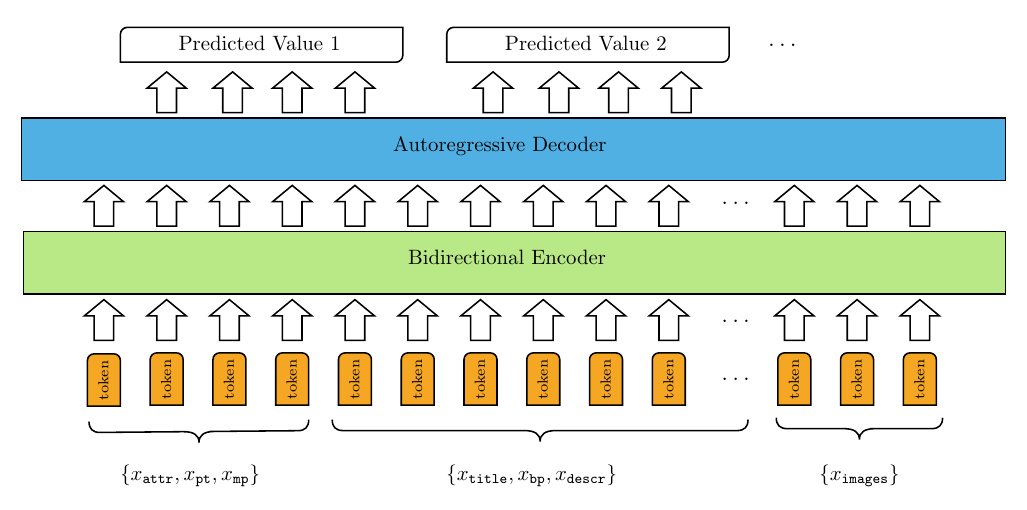}
    \caption{SAGE: Structured Attribute-value Generation}
    \label{fig:sageillustrationnotitle}
\end{figure}

In this work,  we propose modelling $f$ as a \texttt{Seq2Seq} transformer network, with a bidirectional encoder and a left-to-right decoder as shown in Fig.~ \ref{fig:sageillustrationnotitle}. We henceforth refer to the proposed model as \sage.

We fine-tune \sage on input-output pairs of the form
\begin{align}
	\small
	\text{Input: }\{ \texttt{attr},x_\textit{pt},x_\textit{mp}, x_{\textit{title}}, x_{\textit{bullet point}},  x_{\textit{description}} \} & 
	\qquad \text{Output: }\{ v_1 [, v_2, \dots, v_K]\} \nonumber
\end{align}

i.e., \sage is fed inputs containing the relevant information of the \asin and is trained to learn to generate value(s) for the target attribute.

\subsection{Negative Training Labels}
Importantly, we expand set $\mathcal{V}$ with two special values:
\begin{description}
	\item[\text{[\,NA]\,}] Each attribute is applicable to a subset of \asins within each product-type. For example in the product-type SHIRT, the attribute \textit{team\_name} is relevant for a subset of sports shirts, like team jerseys, however, it is surely not applicable for the majority of \asins within the SHIRT product type.  To tackle this challenge, and to help \sage learn to avoid making predictions for irrelevant attributes, we introduce the special value ``[NA]'', which stands for ``Not Applicable''.
	\item[\text{[\,NO]\,}]  For an attribute value to be generated, the related contextual information needs to be present in the input. However, the \asins the model will see in production may not always contain such information. Therefore, training the models only on input-output pairs for which the target attribute value is obtainable does not represent practical application scenario, especially when one fine-tuning over powerful pretrained generative \texttt{Seq2Seq} models. To tackle this challenge, we introduce the special value ``[NO]", which stands for ``Not Obtainable''. 
\end{description}

\subsection{Training Algorithm}
\sage can be fine-tuned over any \texttt{Seq2Seq}/Summarization network. In the experimental section of this work we  assess several architectural options.

To ensure effective training of \sage we follow the classical weak-strong data training methodology. Specifically, we use input-output pairs coming both from the publicly available catalog data, as well as, a small subset of human-verified labeled data. Human-labeled data are high-quality, hence, we consider them as strong labels. On the other hand, public catalog data are plentiful and more diverse; thereby, exposing the model to a rich corpus of text pertaining to products and target attributes. However, catalog data can be noisy, hence are considered as weak data.

\begin{itemize}
	\item[\textbf{Step 1:}]  We initialize the model with pretraining weights and fine-tune it for the attribute-value-prediction task on a combination of weak and strong labels
	\begin{displaymath}
		\left(\mathcal{X}_{\text{train}}^\text{catalog}, \mathcal{V}_\text{train}^\text{catalog} \right) \cup \left(\mathcal{X}_{\text{train}}^\text{human}, \mathcal{V}_\text{train}^\text{human} \right) 
	\end{displaymath}
	for a maximum of 20 epochs, or till the model's performance plateaus, based on the accuracy metric on similarly proportioned eval input-output pairs. 
	\item[\textbf{Step 2:}]  We continue model training starting from the last epoch of step 1 (max 20, or till the model validation accuracy plateaus) and we further fine-tune it using only the strong labels:  
	\begin{displaymath}
		\left(\mathcal{X}_{\text{train}}^\text{human}, \mathcal{V}_\text{train}^\text{human} \right) 
	\end{displaymath}
\end{itemize}

\bigskip

\subsection{Confidence Score Calculation}

During the process of making attribute-value-predictions for the catalog with a machine learning model, it is often important for the model to be able to provide confidence scores for each prediction. Essentially, confidence scores allow for a more fine-tuned control over the quality of the predictions being made by the model and subsequently added to the catalog. To compute the confidence scores associated with \sage predictions we propose utilizing \textit{beam search}.

Beam search is a heuristic search algorithm that explores a graph by expanding the most promising nodes, as determined by an evaluation function relevant to the task at hand. In our context, beam search can be used to generate multiple predicted output sequences, with the top-K such sequences being the ones with the highest probability of being correct, according to the model. The confidence score for the winning sequence can then be obtained by applying a softmax function over these top-K sequences. 

Concretely, if $\{(\texttt{seq}_1,p_1),(\texttt{seq}_2,p_2), \dots, (\texttt{seq}_K,p_K)\}; p_1 \geq p_2 \geq \cdots \geq p_K$, are the output sequences alongside the associated probabilities produced by a \texttt{BeamSearchDecoder} with $K$ beams, by applying a trained \sage model on an input $x$, the final prediction of the model along and the associated confidence score are given by 
\begin{displaymath}
	\{ (\sage(x) , \Pr(\sage(x)))\} \triangleq \left(\texttt{seq}_1, \frac{e^{p_1}}{\sum_{j=1}^{K}e^{p_j}}\right)
\end{displaymath}

Considering large values of $K$ in beam search can lead to better exploration of the space of output sequences, but this improvement comes at a cost of higher inference times. This is a problem for product catalogs that are constantly changing and contain millions of items. Fortunately, we have found that using a value of $K=2$ works well in practice.  Therefore, we set the number of beams to 2 for all the results presented in this paper.

\section{Experimental Setting}

\subsection{Datasets}
For the experimental evaluation of \sage we utilize public data for millions of products pertaining to thousands of PACs across multiple languages. For each PAC we collect \asins and corresponding attribute values from 2 sources: (a) catalog; i.e., attribute values that appear in the catalog, predominantly provided by sellers (b) human-auditors; i.e., attribute values explicitly sourced for the purpose of training and evaluating models. Catalog data are readily available, albeit noisy; Human-labeled data, on the other hand, are highly trusted, however they are expensive to procure and thus, their availability is limited. We henceforth refer to this dataset as \fulldata. Moreover, for our ablation studies we make use of a random subset of \fulldata\ containing 500 randomly selected PACs, which we refer to as \testdata. 
\subsection{Evaluation Methodology and Metrics}
\label{Sec: Training Metrics}
Let $\mathcal{X}$ be a set of asins, and let $\model: \mathcal{X} \to \mathcal{V}\times \mathfrak{R}_{[0,1]}$ be a model which outputs pairs of predictions together with probabilities which quantify the confidence of these prediction. We assume that the model was trained using (\asin, value) pairs 
$(x,v) \in \mathcal{X}_{\text{train}}\times\mathcal{V}_{\text{train}}$, and we denote each generated prediction of the model, as $\model(x)$, and the associated probability for this prediction, as $\Pr(\model(x))$. 
The performance of the model is evaluated on \asins in a set $\mathcal{X}_{\text{test}}$, for which the corresponding ground-truth values $\mathcal{V}_{\text{test}}$ are audited by humans, and which were not used to train the model; i.e.
\begin{math}
	\mathcal{X}_{\text{train}} \cap \mathcal{X}_{\text{test}} = \emptyset .
\end{math} 

We aim to assess the performance of competing models in solving the actual problem of backfilling empty slots in the catalog, accurately, and with the highest possible recall. To ensure that the backfills produced by the model are of high quality we deem a model \textit{acceptable}, only when its predictions that would go into the catalog have precision of at least $P\%$, and, at the same time, the precision estimate itself is of high confidence.
Under this prism, the ideal model would have: a) maximum number of acceptable PACs; i.e., number of PAC-scopes for which the model manages to meet the acceptability criteria, and, b) a maximum recall at precision at-or-above $P\%$ for all acceptable PACs. We refer to these metrics as Acceptance Rate @ P, or $\operatorname{AR@}P$, and $\operatorname{Recall@}P$, and we use them to compare different models' performance. To define these metric concretely, let us specify precisely the process below:
\begin{enumerate}
	\item We apply the model to all $x\in\mathcal{X}_{\text{test}}$ and we get a set of pairs 
	\begin{displaymath}
		\{ (\model(x) , \Pr(\model(x)))\}_{x \in \mathcal{X}_{\text{test}}}
	\end{displaymath} 
	\item We identify the smallest probability threshold $t$ such that the model's precision on the set
	\begin{displaymath}
		\mathcal{A}_{t} = \{ x\in\mathcal{X}_{\text{test}},  \text{ such that } \Pr(\model(x))\geq t,  \}
	\end{displaymath} is at least $P$. 
	If $\mathcal{A}_t$ exists and its cardinality, $S_t = \lvert \mathcal{A}_t \rvert$, is at least $S$, we deem the model \textit{acceptable}, and we estimate its recall at precision at-or-above $P\%$, which we denote $\operatorname{Recall@}P$. The bigger the support, $S$, the better the confidence on the precision estimate. For all results presented in this paper, we set $P = 96$ and $S=30$. 
\end{enumerate}

\section{Experimental Results}

\subsection{Selecting the base architecture}

For selecting the best underlying architecture for our task we conducted the following  experiment. We use \testdata\ and we consider the encoder-decoder networks \mtfive-\texttt{small}, \mtfive-\texttt{base},  \mtfive-\texttt{large}, as well as \mbart\ as the basis for fine-tuning for our task. The results are reported in Table~\ref{table: transformer_architecture_selection}.  

\begin{table}[h!]
	\caption{Selecting the base architecture}
	\label{table: transformer_architecture_selection}
	\centering
	\begin{threeparttable}
		\begin{tabular}{lcc}
			\toprule
			\toprule
			Transformer Architectures & AR@96 & Recall@96 \\
			\midrule
			\myrowcolour
			\mtfive-\texttt{small} \phantom{fdasfasfasdfasdfassvcvxcvxvvafsadfsdfsadsfasdfax} & 82.20\% & 79.14\% \\
			\mtfive-\texttt{base} & 84.64\% & 78.76\% \\
			\myrowcolour
			\mtfive-\texttt{large} & 90.71\% & 81.74\% \\
			\mbart-\texttt{large}  & \highest{94.18\%} & \highest{83.42\%}  \\
			\bottomrule
			\bottomrule
		\end{tabular}
		\begin{tablenotes}
			\item For this experiment we train one model per base architecture on \testdata\ and we report the performance in terms or AR@96 and Recall@96. For accurate comparisons we ensure that exactly the same train-test splits are used across trainings. 
		\end{tablenotes}
	\end{threeparttable}
\end{table}

As is evident from our results \mbart\ manages to perform better than the competing approaches in both metrics of interest. Within the \mtfive\  family we find that the performance on our task increases with the size of the network. This promted us to consider training using \mtfive-\texttt{xl} as well; however, the computational implications of fine-tuning such a large network (and the corresponding inference cost implications),  would render it a suboptimal choice for billion-scale catalogs, even if it could in principle outperform \mbart; thus, we opted not to pursue the \mtfive\ family further.  

\subsection{Assessing the Usefulness of Negative Training Signal}
One of the novel modelling ideas introduced in this research involves the explicit characterization of \textit{non-obtainability} and \textit{non-applicability}, as well as the intentional inclusion of such negative signals during fine-tuning of \sage for the attribute-value-prediction task. Besides the immediate benefit of having a single model that is able to tackle all aspects of Catalog data completeness (i.e., having a model which when does not make a prediction, gives an explicit reason for not doing so); we hypothesize that including such negative signals improves the performance of the model in correctly predicting attribute values even when such values are applicable and humanly obtainable, relying on \asin information the model is also exposed to.    

\begin{table}
	\caption{Effects of including  negative signals during training}
	\label{table: Negative Signal}
	\centering
	\begin{tabular}{lcc}
		\toprule
		\toprule
		Model & AR@96 & Recall@96  \\
		\midrule
		\myrowcolour
		\sage (WITHOUT Negative Signal) \phantom{fdsfasdfasdfasasdfafasdfafdas} & 84.44\% & 81.98\% \\
		\sage (WITH Negative Signal) & \highest{94.18\%} & \highest{83.42\%}\\
		\bottomrule
		\bottomrule
	\end{tabular}
\end{table}
To test this hypothesis, as well as, to quantify the associated boost in performance we  conduct the following experiment. We take \testdata\ and we train two \sage models: one for which ``NA'' and ``NO'' values are used during training, and one for which they are not. The performance of both models is evaluated on human-audited test data ensuring that all \asins that are present in the test set have a ground-truth attribute value, as deemed by human auditors, upon examining the information included in the title, bullet points, and description, for the \asin at hand. In other words, we do not penalize the model that was trained without negative signal, for its inability to predict NA, or NO, given that it had no opportunity to observe such values during training. We see that bringing in NAs and NOs results in a significant absolute boost in terms of AR@96 of +9.74\% as well as an absolute boost in Recall@96 of +1.44\%. 

\vspace*{-0.5em}
\subsection{Performance against Competing Approaches}
We compare the performance of \sage against competing baselines. Our aim is to assess the quality of the novel problem formulation, as well as, the proposed solution we introduced;  using as a yardstick the performance of prior solutions that can tackle this problem at the scale and scope \sage does.
In other words, our aim is not to simply compare the learning capabilities of competing ML approaches; rather, we aim to compare the effectiveness of competing solutions for the problem of attribute-value prediction in world-wide product catalogs. Evaluation of the respective performance relies on exactly the same sets of held-out human-audited attribute values per PAC.

\noindent\textbf{Competing approaches:} The baselines we  consider for this comparison are: 
\begin{itemize}
	\item \texttt{Extraction transformers}: To mitigate the significant costs involved in fine-tuning multiple transformer-based methods on \fulldata, while still faithfully depicting the capabilities of the extraction-based attribute-value-prediction methodologies we start by selecting the best-in-class representative method, which we then use to train on \fulldata\ and test against the rest of the baselines. We considered several state-of-the-art attribute extraction methods, including \citep{EXACT,chiu2016named,yan2021adatag,MAVE},  but we found their performance inadequate for our setting and data. 
	This promted us, to extend the NER formulation of~\citep{yan2021adatag} and build  a multi-valued, multi-lingual broad-scope transformer model pretrained on \atlas  which we fine-tuned for the attribute-value-prediction task on \fulldata. Due to the pronounced class-imbalance problem innate to the NER formulation,  in search of the best possible extraction performance we had to customize the loss function of the model; specifically, upon experimenting extensively with multiple loss functions (including Focal Loss~\citep{lin2017focal}, Dice Loss~\citep{li2019dice}, as well as several region-based losses~\citep{rajaraman2021novel}) we found that a custom compound loss based on Dice and TopK was able to yield the best performance in both AR@96 and Recall@96 on our data.  Fine-tuning  \atlas  on \fulldata, took 25 days on a \texttt{p4.24xlarge} AWS ec2 instance (i.e., on a single-node machine with 8 A-100 GPUs, 100 vCPUs, and 1.1T of RAM).   
	\item \sabex: An ensample of multi-label classifiers (LR, SVM, Random Forests, etc). Each PAC was addressed by a PAC-specific model, trained to predict attribute values from the \asin text within a predefined enumerated list of possible options curated by humans. Let us note here, that despite their simplicity, the restricted scope and the human-curated target labels, makes these models very strong baselines in practice. Upon experimentation, we found that exposing these models to catalog data (i.e., using the weak-strong strategy detailed in Section 2) resulted in a performance decline for the vast majority of the PACs. Thus, we opted to train them utilizing only human-labeled data; which yielded the best results. Training of these models was relatively cheap and trivial to parallelize across PACs. Specifically, training on \fulldata, was completed in less than 22 hours in a single \texttt{c5.9xlarge} machine with 36 vCPUs. 
	\item \sage:  A single \sage model, was trained across all product types, Attributes, and countries in \fulldata, using catalog and human-labeled data as per the methodology detailed in Section 2 of this paper. The training was also performed on a \texttt{p4.24xlarge} instance and was completed in 17 days. 
\end{itemize}

Given that \atlas, and \sabex\ models are restricted to only consume text data, we have similarly restricted \sage to utilize solely text input sources. (We examine the effect of adding also image embeddings to the model in Section~\ref{Sec: Image experiment}).  

Table~\ref{table: main results} reports the performance of the competing methods in terms of AR@96 and Recall@96. \sage clearly outperforms both \atlas\ and \sabex, both in overall percentage
\begin{wraptable}{r}{7.0cm}
	\vspace*{-2em}
	\caption{Performance evaluation against competing methods }
	\label{table: main results}
	\centering
	\small
	\begin{threeparttable}
		\begin{tabular}{lcc}
			\toprule
			\toprule
			Model & AR@96 & Recall@96 \\
			\midrule
			\myrowcolour
			\atlas & 54.27\% & 29.37\%\\ 
			\sabex & 88.62\% & 70.42\%\\ 
			\myrowcolour
			\sage & \highest{95.78\%} & \highest{84.86\%} \\
			\bottomrule
			\bottomrule
		\end{tabular}
	\end{threeparttable}
\end{wraptable}
   of PACs that meet the precision cutoff (i.e., backfill precision at or above 0.96), as well as, in terms of average Recall@96; 
with the differences being statistically significant. Specifically, \sage reaches an AR@96 of\textbf{ 95.78\%} and a Recall@96 of \textbf{84.86\%}, thereby achieving an absolute AR@96 boost of +7.16\%, and Recall@96 boost of +14.44\% over the custom PAC-specific solutions of \sabex.  

Note here that \atlas's recall and precision performance is significantly lower compared to the other two competing approaches. This was expected, and is in accordance with the limitations of extraction-based methods for attribute-value prediction. Indeed, even though the predictions made by this model were highly accurate, the recall of the method across the 627 attributes that were present in \fulldata\ was inconsistent; in cases where the attribute value was explicitly mentioned in the text of the \asin, the model was able to tag it correctly; however, when the attribute value was only implicitly mentioned, or for attributes for which the value relied on common-sense reasoning or defaulting, the performance of the model was considerably lower. 

\begin{table}[h!]
	\caption{Comparison against competing baselines}
	\label{table: attribute-level results}
	\centering
	\begin{adjustbox}{width=1\textwidth}
		\begin{threeparttable}
			\begin{tabular}{lcccccccc}
				\toprule
				\toprule
				& \multicolumn{2}{c}{\it Boolean} &  \multicolumn{2}{c}{\it Numerical} &  \multicolumn{2}{c}{\it Type}  & \multicolumn{2}{c}{\it Material}      \\
				\cmidrule(r){2-3}  \cmidrule(r){4-5}  \cmidrule(r){6-7}   \cmidrule(r){8-9}  
				& AR@96  &  Recall@96 &  AR@96  &  Recall@96 &  AR@96  &  Recall@96  & AR@96  &  Recall@96 \\
				\midrule
				\myrowcolour
				\sabex & 85.71\% & 74.26\%&83.58\% & 55.63\% &87.08\% & 64.98\% &  93.78\% & 77.55\% \\
				\sage\attr & 97.67\% & 95.34\% & 95.51\% & 87.53\% & 96.33\% & 87.72\% & 98.22\% & 80.95\% \\
				\myrowcolour
				\sage & 99.07\% & 93.61\% & 93.69\% & 87.73\% & 96.51\% & 85.40\% & 96.01\% & 78.35\% \\
				\midrule
				& \multicolumn{2}{c}{\it age\_range\_description} &  \multicolumn{2}{c}{\it Style} &  \multicolumn{2}{c}{\it  water\_resistance\_level}  & \multicolumn{2}{c}{\it  control\_method}      \\
				\cmidrule(r){2-3}  \cmidrule(r){4-5}  \cmidrule(r){6-7}   \cmidrule(r){8-9}  
				& AR@96  &  Recall@96 &  AR@96  &  Recall@96 &  AR@96  &  Recall@96  & AR@96  &  Recall@96 \\
				\midrule
				\myrowcolour
				\sabex & 90.00\% & 47.45\%& 86.36\% & 85.34\%& 100.0\% & 95.02\%  & 92.31\% & 81.89\%   \\
				\sage\attr & 97.11\% & 87.56\% & 94.32\% & 81.81\%& 100.0\% & 94.63\% & 94.19\% & 81.21\% \\
				\myrowcolour
				\sage & 97.46\% & 84.98\% & 92.54\% & 80.55\% & 100.0\% & 96.45\% & 95.12\% & 83.10\%\\
				\midrule
				& \multicolumn{2}{c}{\it operation\_mode} &  \multicolumn{2}{c}{\it seasons} &  \multicolumn{2}{c}{\it care\_instructions}  & \multicolumn{2}{c}{\it hardware\_interface}      \\
				\cmidrule(r){2-3}  \cmidrule(r){4-5}  \cmidrule(r){6-7}   \cmidrule(r){8-9}  
				& AR@96  &  Recall@96 &  AR@96  &  Recall@96 &  AR@96  &  Recall@96  & AR@96  &  Recall@96 \\
				\midrule
				\myrowcolour
				\sabex & 100.0\% & 51.58\% & 100.0\% & 52.08\% & 88.89\% & 65.21\% & 100.0\% & 59.96\%   \\
				\sage\attr & 98.55\% & 91.32\%& 100.0\% & 91.23\%& 96.79\% & 90.86\%  & 87.80\% & 80.39\%\\
				\myrowcolour
				\sage & 97.14\% & 88.85\%& 100.0\% & 84.84\%& 96.79\% & 91.42\% & 83.33\% &  75.27\%  \\
				\midrule
				& \multicolumn{2}{c}{\it compatible\_devices} &  \multicolumn{2}{c}{\it target\_gender} &  \multicolumn{2}{c}{\it target\_species}  & \multicolumn{2}{c}{\it form\_factor}      \\
				\cmidrule(r){2-3}  \cmidrule(r){4-5}  \cmidrule(r){6-7}   \cmidrule(r){8-9}  
				& AR@96  &  Recall@96 &  AR@96  &  Recall@96 &  AR@96  &  Recall@96  & AR@96  &  Recall@96 \\
				\midrule
				\myrowcolour
				\sabex & 83.33\% & 67.02\%& 91.89\% & 97.90\%& 100.0\% & 57.41\%  & 90.00\% & 90.50\%  \\
				\sage\attr & 93.43\% & 87.48\% & 99.05\% & 95.29\%  & 96.77\% & 85.32\% & 97.26\% & 80.32\% \\
				\myrowcolour
				\sage & 94.24\% & 82.39\%& 99.04\% & 96.13\% & 100.0\% & 82.73\% & 98.63\% & 81.02\%  \\
				\bottomrule
				\bottomrule
			\end{tabular}
		\end{threeparttable}
	\end{adjustbox}
\end{table}

Table~\ref{table: attribute-level results} presents results for a number of common attributes within \fulldata. Here, for reference, we also include the performance of another variant of \sage, denoted \sage\attr that was trained at attribute-level (i.e., \fulldata\ was covered by a collecting of  Attribute-specific, product type-agnostic, and Country-agnostic \sage models).  Notice that both \sage variants generally perform better than the competing baselines\footnote{Average recall of \sabex\ was better only on 2 attributes, but in both cases the acceptance rate was significantly lower than \sage and \sage\attr.}, with the difference being particularly emphatic for types of attributes for which the sought attribute values, are either missing from the text, or implicit mentioned using periphrastic language. Example attributes of the latter, include \textit{boolean} attributes (e.g., attributes of the form \textit{is\_*}, \textit{has\_*}, etc), \textit{numerical} attributes (e.g., attributes of the form \textit{number\_of*}, \textit{maximum*}, \textit{minimum*}, etc), \textit{age\_range\_description}, \textit{care\_instructions}, among others. 
Performance between \sage variants is comparable for most attributes, with \sage\attr achieving relatively better average recall for certain attributes, like \textit{operation\_mode} and \textit{seasons}.

\subsection{Zero-shot Performance}
\label{sec:zero-shot}
Modern e-Commerce catalogs contain hundreds of thousands of unique PACs. Obtaining training data for all of them would require a monumental effort, as well as, number of human auditors. Therefore solutions that hope to solve the problem of attribute value prediction for billion-scale catalogs, would need to be able to showcase zero-shot capabilities that would allow one to train, a model by collecting human labels for only a subset of the PACs while requiring only eval-labels for the rest. This is precisely the setting we test in this section.

Specifically, we select the largest bucket of related attributes within the  \fulldata\ dataset (i.e., the \textit{Numerical} bucket which comprises attributes such as \textit{item\_weight}, \textit{voltage}, \textit{number\_of\_*} across product types and countries) and we conduct the following experiment. We randomly select 1000 PACs, and we further randomly split them into 2 subsets: 
\begin{itemize}
	\vspace*{-1ex}
	\item \underline{Subset A}: Contains 800 randomly selected PACs. For these we give the model access to human-labeled data during training. 
	\item \underline{Subset B}: Contains the remaining 200 PACs. For these the model is not allowed access to human-labeled data during training. Regarding the model's access to catalog data, we examine two settings: 
	\begin{itemize}
		\item We do not feed the model catalog data for the zero-shot PACs during training. 
		\item We do feed the model catalog data for the zero-shot PACs during training. 
	\end{itemize}
\end{itemize}
We evaluate the performance of the models using held-out human-labeled data across all 1000 PACs.
\begin{table}[h!]
\caption{Assessing the performance of the model on zero-shot predictions}
	\label{table: Zero-shot performance}
	\centering
	\begin{tabular}{lcccc}
		\toprule
		\toprule
		& \multicolumn{2}{c}{\it Supervised Subset} &  \multicolumn{2}{c}{\it Zero-shot Subset}   \\
		\cmidrule(r){2-3}  \cmidrule(r){4-5} 
		Model & AR@96 & Recall@96 & AR@96 & Recall@96  \\
		\midrule
		\myrowcolour
		\sage (WITHOUT catalog-data) & 95.39\% & 80.56\% & 77.90\% & 63.71\%\\
		\sage\attr (WITH catalog-data) & \highest{97.10\%} & \highest{82.42\%} &  \highest{84.53\%} & \highest{74.53\%} \\
		\bottomrule
		\bottomrule
	\end{tabular}
\end{table}

Table~\ref{table: Zero-shot performance} reports the performance of the models in the above setting. 

\sage is able to yield acceptable models in zero-shot PACs under both training data settings. Bringing in catalog data for the zero-shot PACs yields superior performance, with \sage managing to reach an acceptance rate of 84.53\% with a Recall@96 of 74.53\%. The performance of the models in the supervised subset was even better. This was expected and emphasizes the importance of high-quality human-audited data during training. Interestingly, bringing in catalog data for the zero-shot PACs boosted the performance of the model, even in the supervised subset, by more than 1.5\% in both AR@96 and Recall@96.

The results of this experiment suggest that the zero-shot potential of \sage is very promising. Note that we opted to design this experiment in a language-agnostic fashion. Indeed, it would have been an easier task if we were to ensure that zero-shot expansion is done within the same language; rather, we were interested in assessing the zero-shot performance of the model when attribute coherence is the only axis of expansion. Notably, subset B spans 130 unique product types and 26 unique attributes across 10 languages.

\subsection{Attribute Applicability Classification}
Besides attribute value prediction our modeling approach permits \sage to predict whether an attribute is applicable or not for the \asin under consideration. Including NA examples during training allows the model to understand better the attributes, and decide whether attribute values need to be produced or not. \textit{But, how well does the model perform in classifying attributes as applicable or not? }

To tackle this question we perform the following experiment. We randomly sample 1000 PACs, and we collect 500 empty \asins from the catalog for each of these PACs, and we ask human-auditors to assess the applicability of each of the included \asins (i.e., classify the \asin-attribute pair as ``NA'' if they believe that the attribute is not applicable for the \asin; or ``App''  if it is applicable, and also provide the attribute value when possible). Using these data we train a \sage model, as well as,  two custom applicability classifiers: (a) a multilayer neural network binary classifier that is fed \texttt{word2vec} embeddings~\citep{mikolov2013efficient} of the \asin text; and, (b) a custom transformer
\begin{wraptable}{r}{7.8cm}
    \vspace{-2em}
	\caption{Classification accuracy of \sage for the problem of attribute applicability}
	\label{table: Applicability}
	\centering 
	\begin{threeparttable}
		\begin{tabular}{lccc}
			\toprule
			\toprule
			Model &  Overall  &  $>90\%$ &  $\leq90\%$ \\
			\midrule
			\myrowcolour
			\texttt{MLP}  & 72.74\% &  73.30\% &   70.28\%\\
			\texttt{XLMR}& 95.52\% &  96.28\%& 92.18\%  \\
			\myrowcolour
			\sage & \highest{96.87\%} & \highest{97.79\%} & \highest{92.89\%} \\
			\bottomrule
			\bottomrule
		\end{tabular}
		\begin{tablenotes}
			\small
			\item When \sage is able to predict a value for the product, we consider this an ``App'' decision.
		\end{tablenotes}
	\end{threeparttable}
\end{wraptable}
 model (after experimenting with several architectures we found that the \texttt{XLMR} transformer~\citep{xlmr} was able to perform better on this task) fine-tuned for the applicability binary classification task. Table~\ref{table: Applicability} reports the accuracy of all three models.  Column ``Overall'' reports the average accuracy across the 1000 PACs, whereas the rest of the columns report the performance focusing on the buckets for which ground truth applicability was above or below $90\%$, respectively. The results suggest that \sage, even though trained for the more general problem of attribute value prediction, it manages to surpass both custom applicability classifiers in terms of classification accuracy. These results highlight the capability of \sage to successfully tackle the task of attribute applicability; a relevant but often overlooked problem facing world-wide product catalogs.         

\subsection{Multi-modal Attribute Value Prediction}
\label{Sec: Image experiment}
Finally, we close this section by presenting preliminary results from an experiment that expands \sage input with the inclusion of image information. 
\newpage
\begin{wraptable}{r}{8.0cm}
	\caption{\msage\ results }
	\label{table: preliminary mutli-modal results}
	\centering
	\begin{threeparttable}
		\begin{tabular}{lcc}
			\toprule
			\toprule
			Model & AR@96 & Recall@96 \\
			\midrule
			\myrowcolour
			\sage (text) & 94.18\% & 83.42\% \\
			\msage (text + images) & \highest{97.19\%} & \highest{85.48\%} \\
			\bottomrule
			\bottomrule
		\end{tabular}
	\end{threeparttable}
\end{wraptable}
To assess the usefulness of adding images to our model, we conduct the following experiment: For all PACs included we compute image embeddings of the MAIN image based on~\citep{Zhao2019}, and we add these embeddings as an input to the encoder of our model, effectively enforcing \textit{early fusion} of the modalities of the \asin input. We denote the corresponding \sage variant as \msage, and we initialize the model with the weights of the corresponding fine-tuned text-only \sage model which we also include for comparison. The results are presented in Table~\ref{table: preliminary mutli-modal results}. Even though for simplicity, \msage was limited to only using general-purpose image embeddings of the MAIN image (arguably, not always the most informative image for every sought attribute value) it manages to outperform the text-only variant yielding more acceptable PACs (+3\%), as well as, better average Recall@96 (+2\%), compared to its text-only counterpart.

\section{Conclusions}
Structured attributes play a crucial role in the product exploration process for customers, who use them to search for, browse, compare, and ultimately decide which products to purchase. Missing product metadata can hinder product discovery, as it can prevent items from being properly organized into refinements, variation families, and comparison tables, leaving customers with insufficient information to make informed purchase decisions. 

In this work, we tackle this problem by introducing a novel problem formulation for the attribute-value-prediction task that overcomes the limitations of current state-of-the-art extraction-based methodologies. We introduce \sage; the first multilingual, Seq2Seq transformer model for attribute value generation. The results of our experiments demonstrate that \sage offers a more comprehensive and effective solution to the attribute-value-prediction problem, paving the path towards improved methodologies that can lead to improvements in the quality and completeness of online e-commerce product catalogs.

\end{document}